\documentstyle[12pt]{article}
\begin{document}
\title {Evaluation of the mean intensity of the P--odd mixing 
of nuclear compound states}
\author {V.A.Rodin and M.H.Urin}
\date{\today}
\maketitle
\begin{center}
{\it Moscow Engineering Physics Institute, 115409, Moscow, Russia}
\end{center}
\begin{abstract}
A temperature version of the shell-optical-model approach for
describing the low-energy compound-to-compound transitions
induced by external single-particle fields is given.  The
approach is applied to evaluate the mean intensity of the P-odd
mixing of nuclear compound states. Unified description for the
mixing and electromagnetic transitions allows one to evaluate
the mean intensity without the use of free parameters.  The
valence-mechanism contribution to the mentioned intensity is
also evaluated.  Calculation results are compared with the data
deduced from cross sections of relevant neutron-induced
reactions.
\end{abstract}

{\bf 1.} The weak nucleon-nucleon interaction gives rise to the
parity-violating part $V_w$ of the nuclear mean field.  In the
case of intermediate and heavy mass nuclei, this part is the
main source of the P-odd mixing of compound states (CS)
corresponding, as a rule, to neutron resonances.  In view of
complexity of CS wave functions only the mean intensity of the
mixing seems to be a subject for theoretical studies. This
intensity is usually characterized either by root-mean-squared
matrix elements $M_w$ of the field $\hat V_w=\sum_a V_w(a)$
between the CS wave functions or by the quantity
$\Gamma^{\downarrow}_w=2\pi M^2_w/d_c$ ($d^{-1}_c$ is the
density of CS).  The experimental values of the weak matrix
elements $M_w^{exp}$ are deduced from the cross sections of the
scattering of the low-energy polarized neutrons from nuclei.
These matrix elements are known now for the $^{233}Th$,
$^{239}U$  \cite{bow} and for $^{114}Cd$, $^{116}In$ \cite{shar}
compound nuclei.  The weak matrix elements found for some other
nuclei are usually assigned to the mixing of certain $p_{1/2}$
and $s_{1/2}$ CS (see, e.g., ref.\cite{krup}).

Some experimental data on the intensities of electromagnetic
low-energy compound-to-compound ($c-c'$) transitions are also
available.  These intensities are characterized by the photon
strength functions $s_\gamma(\omega)=\Gamma_\gamma (\omega)/d_c$
($\omega$ is the transition energy, $\Gamma_\gamma (\omega)$ is
the mean partial radiative width ). The mean total radiative
width $\Gamma^{tot}_{\gamma}$ of neutron resonances  is the
integral characteristic of the mentioned transitions.  The
experimental data on $\Gamma^{tot}_{\gamma}$ are available for
many nuclei \cite{mug}.

Being analyzed in terms of the mean quantities
$\Gamma^{\downarrow}_w$ the $c-c'$ P-odd mixing can be
considered as the weak $c-c'$ transitions (with $\omega=0$) by
analogy with the low-energy electromagnetic $c-c'$ transitions.
Really, both of the transitions are induced by the
single-particle fields ($\hat V_w$ and $\hat V_{E1,M1}$,
respectively) and, therefore, take place only due to coupling of
particle-hole and many-quasiparticle configurations.  For these
reasons, the transitions can be described in terms of the
particle-hole (or giant-resonance) strength functions taken in
the low-energy limit, i.e. for the energy $\omega$, which is less
than the energy $D$ of the corresponding giant resonance ($0^-$
and $1^-,1^+$, respectively). Thus, the unified description for
the weak and electromagnetic low-energy $c-c'$ transitions seems
to be profitable because the same adjustable parameter is used
for calculations of the corresponding strength functions (see
below).

 Since the first qualitative estimations of mean intensities of
$c-c'$ transitions \cite{blin}, two alternative mechanisms
(valence and statistical) are widely discussed
\cite{zar}--\cite{flam}. According to the valence mechanism, the
$c-c'$ transitions are due to single-quasiparticle components of
the CS wave functions, whereas according to the statistical
mechanism, they are due to many-quasiparticle components. Most
of the authors  agree in that the statistical (or temperature)
mechanism is dominant (see, however, ref.\cite{zar}). The
valence mechanism can provide a special interest \cite{ur2}.
Note in this connection that the method for conversion from the
product of the mean neutron widths of the $s_{1/2}$ and
$p_{1/2}$ compound resonances to the valence contribution to the
$M^2_w$ value is given in refs.\cite{ur1,ur2}. The other
considerations of this contribution \cite{zar,flam} seem to be
only estimative.

It is known, that the photon strength function
$s_\gamma(\omega,U_{c'})$ is proportional to the energy-averaged
total cross section of photoabsorption by the nucleus being in
the final state. For this reason, it is convenient to use the
following definition of strength function $S_V(\omega,U_{c'})$
for the $c-c'$ transitions induced by an external field $\hat V$
: $S_V(\omega,U_{c'})= <\sum_{c}\mid \hat V_{cc'}\mid ^2
\delta (U_{c}-U_{c'}-\omega)>$ , where $U_{c}$ and $U_{c'}$ are
excitation energies of initial ($c$) and final ($c'$) nuclear
states, respectively; brackets $<\dots >$ are taken to mean
averaging over energy interval $I \gg d_c$ .  Bearing in mind
averaging over final CS we can consider the strength function
$S_V(\omega,U')=M^2_V/d_c$ as the quantity, which is calculated
for the nucleus heated up to the temperature $t(U')$
corresponding to the mean energy $U'$ of final CS
($U'=U-\omega$, $U$ is the mean energy of initial CS). The idea
to evaluate $S_V(\omega,U')$ by this way seems to be adequate to
the considered problem, but the method for practical realization
of this idea is another matter.  The method for description of
the particle-hole excitations corresponding to giant resonances
in heated nuclei has long been used and consists in the
temperature modification of nucleon occupation numbers in the
RPA equations (see, e.g., ref.\cite{ign}). But there are no
well-grounded method for the description of giant-resonance
spreading due to the coupling of the particle-hole states to
many-quasiparticle configurations in heated nuclei. As mentioned
above the low-energy $c-c'$ transitions induced by a
single-particle field take place only due to this coupling. In
other words, in heated nuclei the low-energy (up to $\omega=0$)
''tail'' of giant resonances arises due to the mentioned
coupling. This statement has been widely used in attempts to
evaluate the ($c-c'$)-transition intensities by means of an
extrapolation of the Lorentz--like parametrization of the
giant-resonance strength functions to small $\omega$
\cite{kad,auer}. The approach used in refs.\cite{ur1} (called by
the authors as semiempirical one) and based on both
microscopical (shell-model) description of particle-hole
configurations and a phenomenological consideration of
quasiparticle damping (in terms of the imaginary part of an
optical potential) seems to be the most advanced description for
the ''tail'' of giant resonances in heated nuclei. The $M_w$
evaluation based on a consideration of the so-called
''principal'' components of the CS wave functions \cite{flam}
and on the microscopical description of many-particle states
\cite{bow1} were realized only for $^{233}Th$ and $^{239}U$
compound nuclei, respectively.

In this work the approach used in refs.\cite{ur1} for evaluation
of $M_w$ is extended in the following lines.  (i) We take into
consideration the complementary term in the expression for the
strength function $S_V(\omega,U')$. This term (omitted in
ref.\cite{ur1}) ensures the $\omega \to -\omega$ invariance of
this strength function and gives essential contribution to $S_V$
for small $\omega$.  (ii) For each considered nucleus we find
the value of single adjustable parameter (describing the
strength of particle-hole coupling to many-quasiparticle
configurations) by fitting the calculated value of
$\Gamma^{tot}_\gamma$ to the experimental one. It allows us to
calculate the root-mean-squared matrix element $M_w$ without the
use of free parameters because other model parameters can be
chosen from independent data.(iii) We evaluate $M_w$ for many
nuclei (with $A>50$) and compare the calculation results with
available experimental data.(iv) For all considered nuclei we
evaluate also the valence-mechanism contribution to $M_w$ by the
convenient method proposed in ref.\cite{ur2}.

{\bf 2.} The semiempirical approach for calculating the
low-energy part of the strength function $S_V(\omega,U')$
corresponding to a single-particle external field $V$ is
formulated in two steps. The first one is formulated for cold
nuclei ($U'=0$). The method for calculating the giant-resonance
strength function $S_V(\omega \sim D,0)$ has long been
formulated in following approximations \cite{dov}: (i) the
shell-model basis of appropriate particle-hole states has been
limited by discrete (and quasidiscrete) spectrum; (ii) spreading
of single-particle and single-hole states has been described in
terms of the imaginary part of an optical-model potential. Thus,
the method can be called as the shell-optical-model approach.
For small excitation energies ($\omega \ll D$) the approach has
been generalized to the case of exact consideration for the
single-particle continuum \cite{ur3} by the methods of the
finite Fermi-system theory \cite{mig}. The shortcoming of the
approaches developed in refs.\cite{dov,ur3} is neglect of
possible interference of particle and hole damping at the
description of the giant-resonance spreading. For this reason
the imaginary part of an optical potential used in these
approaches should be considered as effective one.

We give the basic equations of the shell-optical-model approach
in rather schematic form:
\begin{equation}
S_V(\omega,0)=-{1\over{\pi}} Im \int \tilde V^{+}(x,\omega)
A(x,x';\omega)\tilde V(x',\omega)dxdx' . 
\end{equation}
\begin{equation}
\tilde V(x,\omega)= V(x)+\int F(x,x_1)
A(x_1,x_2;\omega)\tilde V(x_2,\omega)dx_1dx_2 . 
\end{equation}
\noindent Here $x$ is the set of nucleon coordinates including
spin and isospin variables, ${\tilde V}(x,\omega)$ is the
effective field that differs from the external field $V(x)$ due
to the polarization effect caused by the particle-hole
interaction $F(x,x')$; $A(x,x';\omega)$ is the energy-averaged
particle-hole propagator, which is the main quantity in the
considered problem. Within the shell-model approach, when the
coupling of particle-hole and many-quasiparticle configurations
is neglected, eqs.(1),(2) correspond to the continuum-RPA,
widely used in the theory of giant resonances
\cite{shlber,rumur}.  In this case $A \to A^{(RPA)}$, where
\begin{equation}
A^{(RPA)}(x,x';\omega)=\sum_{\nu}
n_{\nu}\phi_{\nu}^{*}(x)\phi_{\nu}(x')
(g^{(0)}(x,x';\varepsilon_\nu+\omega)+
g^{(0)}(x,x';\varepsilon_\nu-\omega)),
\end{equation}
\noindent Here $\varepsilon_\nu$ and $\phi_{\nu}$ are the
single-particle energies and  bound-state wave functions,
respectively; $n_{\nu}$ are occupation numbers;
$g^{(0)}(x,x';\varepsilon)$ is the Green function of the
single-particle Schr\"odinger equation with the shell-model
potential $u(x)$. In the low-energy limit $\omega \ll D$ the
function $\int A^{(RPA)}(x,x';\omega) {\tilde V}(x',\omega)dx'$
has no poles in the complex $\omega$-plane, so that
$S_V^{(RPA)}(\omega \ll D,0)=0$.

A non-zero value of $S_V(\omega \ll D,0)$ results from the
coupling of particle-hole states to many-quasiparticle
configurations. The consideration for this coupling is beyond of
the RPA and has been undertaken within the shell-optical-model
approach. The expression for $A(x,x';\omega)$ has been obtained
in ref.\cite{ur3} with the use of the spectral expansion for
single-particle Green function in the coordinate representation:

\begin{equation}
A\simeq A^{(RPA)}+i~ImA;\ 
ImA(x,x';\omega)= Im(\delta A(x,x';\omega)+
\delta A(x,x';-\omega)), 
\end{equation}
\begin{equation} Im\delta A(x,x';\omega)\simeq 
Im\sum_{\nu}\phi_{\nu}^{*}(x)
\phi_{\nu}(x')(n_{\nu}g^{(p)}(x,x';\varepsilon_\nu+
\omega)-(1-n_{\nu})g^{(h)}(x,x';\varepsilon_\nu-\omega)).
\end{equation}

\noindent Here $g^{(p,h)}(x,x';\varepsilon)$ is the Green
function of the single-particle Schr\"odinger equation with the
optical-model potential
$u^{opt}(x)=u(x)-iw^{(p,h)}(\varepsilon)f(r)$, where the
intensity of the imaginary part of the optical potential for
particles $(p)$ and holes $(h)$ is defined as follows ($f(r)$ is
the formfactor of this part):

\begin{equation}
w^{(p)}(\varepsilon)=w(\mid \varepsilon-\mu \mid)
(1-\theta(\varepsilon-\mu))\ ; \
w^{(h)}(\varepsilon)=-w(\mid \varepsilon-\mu \mid)
\theta (\varepsilon-\mu). 
\end{equation}

\noindent Here $\mu$ is the the chemical potential determined by
the equation $\sum_\nu n_{\nu}=N$, where $N$ is the number of
particles for the given subsystem; $\theta(y)=1$, when $y<0$,
$\theta(y)=0$, when $y>0$.  In the low-energy limit $\mid \int
A^{(RPA)}(x,x';\omega) {\tilde V}(x',\omega)dx' \mid \gg \mid
\int ImA(x,x';\omega) {\tilde V}(x',\omega)dx'\mid$ (as it
follows from eqs.(3)-(6) when $\omega \ll D$ and $\mid w \mid
\ll D$), so that the quantity $ A^{(RPA)} $ can be used in
eq.(2) for the effective (quasistatic) field. To clarify the
meaning of eqs.(4)-(6) we notice, that (i) the first and the
second terms in eq.(5) correspond to damping (or spreading)
single-particle and single-hole states, respectively; (ii)
$Im\delta A(\omega \to 0)\to 0$, because the quasiparticle
damping vanishes provided that the quasiparticle energy $\mid
\varepsilon-\mu \mid$ tends to zero (for instance,
$w^{(p,h)}(\varepsilon)=\alpha(\varepsilon-\mu)^2$  \cite{mig});
(iii) $Im\delta A(-\omega)$ is identically equal to zero
($\omega \ge 0$), because in cold nuclei there are no both
single-particle excitations with $\varepsilon<\mu$ and
single-hole excitations with $\varepsilon>\mu$.

At first sight the above expressions can be directly applied for
calculating the particle-hole strength function $S_V(\omega,U')$
in accordance with the known Brink's hypothesis. That is correct
provided that $\omega \gg t(U')$. The last statement has been
checked by calculations of the giant-resonance strength function
for heated nuclei ($\omega \sim D, \omega \gg t$) \cite{ign}.
However, it is obvious, that the Brink's hypothesis should be
modified for the case $\omega \sim t$ because of temperature
smearing out the Fermi surface.

A semiempirical way to take this effect into consideration has
been proposed in ref.\cite{ur3} and partially realized in
refs.\cite{ur1}. The basic point of the semiempirical approach
(and the second step in its formulation) consists in the
reasonable assumption that one can come to the case of heated
nuclei by the {\it a priori} use of an appropriate modification
of the occupation numbers and intensity $w(\mid\varepsilon-\mu
\mid)$ in eqs.(4)-(6):

\begin{eqnarray}
& \theta(\varepsilon-\mu)\longrightarrow\theta_t(\varepsilon-\mu)
\equiv (1+exp[(\varepsilon-\mu)/t])^{-1} ,\nonumber \\
& n_{\nu}\longrightarrow\theta_t(\varepsilon_{\nu}-\mu) ,
\hskip1.5cm 
\sum_{\nu}\theta_t(\varepsilon_{\nu}-\mu)=N , \\
& w(\mid \varepsilon-\mu \mid)\longrightarrow
w(\mid \varepsilon-\mu \mid,t)=\alpha((\varepsilon-\mu)^2+
(\pi t)^2).
\nonumber
\end{eqnarray}

\noindent These equations are similar to those that are used to
describe the single-quasiparticle damping in infinite
Fermi-systems at the finite temperature \cite{pin}.  The first
two equations describe temperature smearing out the Fermi
surface in heated nuclei. Due to this effect: (i) both
single-particle excitations with $\varepsilon<\mu$ and
single-hole excitations with $\varepsilon>\mu$ are possible;
(ii) the intensity $w(\mid \varepsilon-\mu \mid)$ is modified
and $w(\varepsilon \to \mu,t) \ne 0$ as opposite to the case of
cold nuclei. As consequences of the above statements two results
of the substitution of eqs.(7) into eqs.(1),(4)-(6) should be
noted. The quantity $Im\delta A(-\omega)$ in eq.(4) is
comparable with $Im\delta A(\omega)$ up to $\omega\simeq t$
(taking this quantity into account ensures the $\omega \to
-\omega$ invariance of strength functions $S_V(\omega,U')$). In
the static limit ($\omega=0$) the value of the strength function
$S_V(0,U)$ is nonzero, as it follows from eqs.(1)-(7), and can
be estimated as $\alpha (\pi t D^{-1} M_V^{sp})^2 A^{1/3}$,
where $M_V^{sp}$ is the so-called single-particle matrix
element. The last quantity is not well-defined one when the
low-energy transitions ($\omega \ll D$) are studied.
Nevertheless, it is considered in most attempts to estimate the
$M_V$ value \cite{zar,kad,flam}. The use of the coordinate
representation for the particle-hole propagator (4),(5) allows
one to avoid the use of $M_V^{sp}$ because the whole
single-particle spectrum is exactly taken into account. Similar
statements about the use of $M_V^{sp}$ are valid in
consideration of the valence-mechanism contribution to the $M_V$
value (see Sect.4 and compare refs.\cite{zar,flam} with
refs.\cite{ur1,ur2}).

The temperature $t$ in eqs.(7) is naturally determined by the
mean excitation energy of the final compound states:
$t=\sqrt{U'/a}$, where $a$ is the parameter in the well-known
semiempirical formula for the level density (see e.g.
ref.\cite{dil}). One can suppose that possible errors in the
calculation of the strength function $S_V(\omega,U')$, which are
connected with the {\it a priori} use of eqs.(7) in
eqs.(1),(4)-(6), can be compensated, to a certain extent, by an
appropriate choice of single adjustable parameter $\alpha$ in
the expression for $w(\mid\varepsilon-\mu \mid,t)$.  For
instance, if one finds this parameter by fitting the calculated
value of $\Gamma^{tot}_\gamma$ to the experimental one for the
given nucleus, then one can calculate the
$\Gamma^{\downarrow}_w$ and $M_w^2$ values for the same nucleus
without the use of free parameters.  Such a procedure is
realized below.

{\bf 3.} The mean total radiative width of neutron resonances
can be calculated by means of the well-known expression
(see,e.g.  ref.\cite{Gamma}):

\begin{eqnarray}
& \Gamma^{tot}_{\gamma}=\Gamma^{tot}_{\gamma}(E1)+
\Gamma^{tot}_{\gamma}(M1),\nonumber \\
& \Gamma^{tot}_{\gamma}(E1,M1)=3\rho^{-1}_0(U_n)\int^{U_n}_0 
s_{E1,M1}(\omega,U_n-\omega)\rho_0(U_n-{\omega})d\omega,
\end{eqnarray}

\noindent where $s_{E1,M1}$ are the photon strength functions
mentioned above,$U_n=B_n-2\Delta$, $B_n$ is the neutron binding
energy, $\Delta$ is the adjustable parameter in the
level-density formula \cite{dil} (for double-even nuclei
$\Delta$ is approximately equal to the pairing energy per
nucleon), $\rho_0(U)\sim (aU^2)^{-1}\exp (2\sqrt{aU})$ is the
density of levels with the zero angular momentum. The external
fields corresponding to E1 and M1 transitions can be chosen in
the form:

\begin{equation}
V_{E1}(x)=-\tau^{(3)}z,\hskip.5cm
V_{M1}(x)={1\over 2}\{ \mu_{n}(1+\tau^{(3)})+(\mu_p-1/2)(1-\tau^{(3)})\} 
\sigma^{(3)}, 
\end{equation}

\noindent where $\tau^{(3)}$ and $\sigma^{(3)}$ are the isospin
and spin Pauli matrixes respectively; $z$ is the nucleon
coordinate ; $\mu_{n}=-1.91$, $\mu_{p}=2.79$. According to the
well-known expressions for the dipole partial widths
$\Gamma_\gamma (\omega)$, the photon strength functions
$s_{E1,M1}$ are determined by the strength functions $S_{E1,M1}$
corresponding to fields (9) as follows:

\begin{equation}
s_{E1,M1}(\omega,U)={\omega^3 \over 3k_{E1,M1}}S_{E1,M1}(\omega,U), 
\end{equation}

\noindent where $k_{E1}=137\cdot(\hbar c)^2$,
$k_{M1}=137\cdot (mc^2)^2$, $m$ is the nucleon mass.

The weak nuclear mean field is chosen in the form \cite{des}:

\begin{equation}
V_w={\pi \over{2}}10^{-6}\cdot \Lambda^3_N c
\cdot \{ \xi_{n}(1+\tau^{(3)})+
\xi_p(1-\tau^{(3)}) \} [\vec {\sigma}\vec {p},\varrho(r)]_+ ,
\end{equation}

\noindent where $\Lambda_N=\hbar /mc$, $\vec {p}$ is the nucleon
momentum, $\varrho$ is the nuclear density normalized to the
total number of nucleons $A$. Calculations were performed using
$\varrho(r)={3\over{4\pi}}r_0^{-3}f_{WS}(r,R,a_0)$, where
$f_{WS}$ is the Woods--Saxon function, $r_0=1.24\ fm$,
$R=r_0A^{1/3}$, $a_0=0.65\ fm$, and two sets of the parameters of
the weak nucleon-nucleus interaction \cite{des}:
$\xi_n=-0.7$,$\xi_p=3.3$ (set (I)) and $\xi_n=\xi_p=3.3$
(set (II)). Then the values of $M^2_w$ and $\Gamma^{\downarrow}_w$
are determined by the strength function corresponding to field (11):

\begin{equation}
M^2_w=S_w(\omega=0,U=U_n)d_c\ ;
\ \Gamma^{\downarrow}_w=2\pi S_w(\omega=0,U=U_n)  
\end{equation}

\noindent Using the above-given estimation of $S_V(0,U)$ with 
$D \simeq \varepsilon_FA^{-1/3}$, $\alpha \simeq
0.1 MeV^{-1}$, $t \simeq 0.5 MeV$ ($\varepsilon_F\simeq 
40 MeV$ is the Fermi energy) and the
amplitude of field (11) with $\xi=1$, $p\simeq p_F=\sqrt
{2m\varepsilon _F}$, we obtain $S_w \sim 10^{-8}eV=10 neV$.

Except for the parameter $\alpha$ in eqs.(7), two types of input
data are used for evaluating $M^2_V$  within the framework of
the temperature version of the shell-optical-model approach.
The first one is the nuclear mean field and the particle-hole
interaction. In the following we use the Landau-Migdal
particle-hole interaction \cite{mig}:

\begin{equation}
F(x_1,x_2)=C\{ f+f'({\vec {\tau}}_1{\vec {\tau}}_2)+
(g+g'({\vec {\tau}}_1{\vec {\tau}}_2))
({\vec {\sigma}}_1{\vec {\sigma}}_2) \} 
\delta({\vec r}_1-{\vec r}_2), 
\end{equation}

\noindent where the dimensionless phenomenological parameters
$f',g,g'$ are involved into the calculation of the strength
functions $S_V$ corresponding to the external fields (9) (the
strength parameter $C=300\ MeV\cdot fm^3$). The mentioned
parameters are chosen as follows:$f'=0.88\cdot (1+2.55\cdot
A^{-2/3})$, $g=g'=0.7$ . As for the nuclear mean field, we use
the shell-model potential of the Saxon-Woods type from
ref.\cite{chep}. The set of parameters in the semiempirical
formula for the level density is the second type of input data
for the calculations. These parameters are taken from
ref.\cite{dil}.

For practical calculations of strength functions $S_V$ it is
necessary to separate the spin-angular and isospin variables in
basic eqs.(1),(2), in which the external fields (9),(11) are
used. This procedure is straightforward and leads to the rather
cumbersome final formulae. For brevity sake, we give here the
corresponding formulae only for the case of weak $c-c'$
transitions in the form, which is rather different from that
given in the second reference \cite{ur1}. In the case of E1
transitions the corresponding formulae are just the same except
for taking into account term $\delta A(x,x';-\omega)$ in eq.(4).
According to eqs. (1),(4)-(7),(11) we get the following
expressions for $S_w$:

\begin{eqnarray}
& S_w= \sum_{\beta=n,p}
s^{\beta}_w \cdot \xi_\beta^2 ,\nonumber \\
& s^{\beta}_w=
{2\over{\pi}} e_w^2 Im \int drdr' \sum_{\lambda\lambda'}
t_{(\lambda)(\lambda')}\chi_{\lambda}^{\beta}(r) 
{\hat L}^{r'}_{(\lambda')(\lambda)}[\chi_{\lambda}^{\beta}(r')] 
{\hat L}^{r}_{(\lambda)(\lambda')}
[\theta_t(\varepsilon_\lambda^{\beta}-\mu^{\beta})
g^{(p)\beta}_{(\lambda')}(r,r';\varepsilon_\lambda^\beta)  
\nonumber \\
& -(1-\theta_t(\varepsilon_\lambda^{\beta}-\mu^{\beta}))
g^{(h)\beta}_{(\lambda')}(r,r';\varepsilon_\lambda^\beta)] 
\end{eqnarray}

\noindent Here $e_w=0.75\cdot 10^{-6}(\Lambda_N / r_0)^3
\cdot \hbar c$; $\varepsilon_\lambda^{n (p)}$ and 
$\chi_{\lambda}^{n (p)}(r)$ are, respectively, the energy and
the radial wave functions for the single-neutron(proton) bound
states determined by the (real) shell-model potential;
$\lambda=\varepsilon_\lambda,j_\lambda,l_\lambda \equiv
\varepsilon_\lambda,(\lambda)$ is the set of the bound-state
quantum numbers; $g_{(\lambda)}(r,r';\varepsilon)$ is the Green
function of the Schr\"odinger radial equation with the
optical-model potential whose real part is coincident with the
shell-model potential and the imaginary part is defined for
particles and holes according to eqs.(6),(7);
$t_{(\lambda)(\lambda')}=(2j_\lambda+1)\delta_{j_\lambda
j_{\lambda'}}$; the operator ${\hat L}^r_{(\lambda)(\lambda')}$
is defined as follows:

$${\hat L}^r_{(\lambda)(\lambda')}=2f_{WS}(r,R,a_0)({\partial 
\over{\partial r}}+{B_{(\lambda)(\lambda')}\over{r}})+
{\partial f_{WS}(r,R,a_0)\over{\partial r}}, $$

\noindent where $B_{(\lambda')(\lambda)}=
(l_\lambda(l_\lambda+1)-l_{\lambda '}(l_{\lambda '}+1))/2$.  In
the case of the static ($\omega=0$) weak transitions there is no
core-polarization effect caused by the momentum-independent
interaction (13). Really, because of time-reversal invariance of
the field $V_w$ (11) only the additional term proportional to
$\omega \vec {\sigma} \vec {r}/r$ can appear in the expression
for the effective field $\tilde V_w$ satisfying eq.(2). The
static core-polarization effect takes place only due to the
momentum-dependent part of the particle-hole interaction.  One
can expect that the intensity of this part is small \cite{mig}.

As mentioned above, the single adjustable parameter $\alpha$
determining the strength of the optical-potential imaginary part
according to eqs.(7) can be found for each nucleus by fitting
the calculated value of $\Gamma^{tot}_\gamma$ to the
experimental one. The experimental values of
$\Gamma^{tot}_{\gamma}$ are taken from ref.\cite{mug}. The
calculations are performed according to eqs.(1)-(10). The radial
Green functions are calculated by means of the regular and
nonregular solutions of the corresponding Schr\"odinger
equation. The formfactor $f(r)$ of $w(r,\varepsilon)$ is taken
in the volume form ($f(r)=f_{WS}(r,R,a_0)$).  The parameters
needed for the calculation of both the level density and
temperature are taken from ref.\cite{dil}. The results of
$\alpha$ matching for several nuclei are given in the Table I.
The calculated values of relative contribution of M1 transitions
to $\Gamma^{tot}_{\gamma}$ fall into the range $(0.05-0.35)$. In
some cases this contribution is relatively large due to the
existence of the low-energy single-particle M1 transitions,
which has been taken into account in the $S_{M1}$ calculation by
the method given in ref.\cite{dov}. The accuracy of the
$\Gamma^{tot}_{\gamma}$ calculation and, therefore, of $\alpha$
matching is not high mainly due to: (i) the uncertainties in the
level-density parameters $a$ and $\Delta$ (see e.g.
ref.\cite{dil}); (ii) neglect of the contribution of the E1
transitions to the ground and low-lying states of simple nature
to $\Gamma^{tot}_{\gamma}$; (iii) neglect of the nucleon pairing
at finite temperature. Notice in this connection that it is not
necessary to match $\alpha$ with a high accuracy because the
matrix elements $M_w$ are proportinal to $\alpha^{1/2}$.

The $\alpha$ parameters found by the presented way are further
used for calculating $S_{w}$ according to eq.(14). The results
of the $s_{w}^{n,p}$ and $M_w$ calculations are also given in
the Table I. Notice, that the calculations are performed for the
$^{233}Th$ and $^{239}U$ compound nuclei as for spherical ones.
The mean energy interval $d_s$ between $s_{1/2}$ compound
resonances at the neutron binding energy for the considered
nuclei is taken from ref.\cite{mug}.  Bearing in mind that in
some cases the calculated root-mean-squared matrix elements
$M_w$ are compared with the ''individual'' experimental matrix
elements, the agreement with the known $M_w^{exp}$ values seems
to be satisfactory (see the Table I).  Notice also, that in the
case of even--$A$ compound nuclei the $M_w^{exp}$ values given
in the Table I are actually only the low limit of $M_w^{exp}$
\cite{pik}.

{\bf 4.} Let us turn to calculation of the valence part of the
$M^2_w$ value following refs.\cite{ur1,ur2}. As shown in
ref.\cite{ur1} the quantity $(M^2_w)_{val}$ can be calculated by
means of an optical model. The more convenient way has been
proposed in ref.\cite{ur2} where the method of conversion of the
reduced (to $1 eV$) mean neutron widths $\Gamma^0_{s_{1/2}}$ and
$\Gamma^0_{p_{1/2}}$ of $s_{1/2}$ and $p_{1/2}$ compound
resonances to the $(M^2_w)_{val}$ value was given. After
separation of spin--angular variables in the corresponding
expression obtained in ref.\cite{ur2} we get

\begin{eqnarray}
&(M^2_w)_{val}=\Gamma^0_{s_{1/2}} \Gamma^0_{p_{1/2}}
 q^2  & \ \nonumber \\
 \\
&q^2=-({\varepsilon \over {1 eV}})^2 
{2 \over {\pi^2}} { \xi_n^2 e_w^2 \over{S_{s_{1/2}} S_{p_{1/2}}}}
\int drdr' {\hat L}^{r}_{(p_{1/2})(s_{1/2})}
[Im g^{n}_{(s_{1/2})}(r,r';\varepsilon)]
{\hat L}^{r'}_{(s_{1/2})(p_{1/2})}
[Im g^{n}_{(p_{1/2})}(r,r';\varepsilon)] & \ \nonumber
\end{eqnarray}

\noindent Here $S_{s_{1/2},p_{1/2}}$ are the $s_{1/2}$ and
$p_{1/2}$ neutron strength functions, $\varepsilon$ is the
neutron kinetic energy.

The following comments to exprs.(15) should be given. (i)
Because the neutron strength functions can be calculated by
means of the optical model ($S_{s_{1/2},p_{1/2}}={2 \over {\pi}}
\eta_{s_{1/2},p_{1/2}}$, where $\eta_{s,p}$ are the imaginary
part of the neutron scattering phases), the quantity $q^2$ can
be also calculated within the optical model.  (ii) In contrast to
its physical meaning, the quantity $(M^2_w)_{val}$ determined by
eqs.(15) is not equal to zero when $w=0$ due to the contribution
of single-particle continuum to $Im g(r,r';\varepsilon >0)$. To
exclude this contribution the following substitution should be
used in eqs.(15) (compare with consideration of the valence part
of the E1-radiative strength function of neutron resonances
\cite{ur4}):

\begin{equation}
Img(r,r';\varepsilon) \to Img(r,r';\varepsilon)+\pi 
\chi_\varepsilon^{opt}(r)\chi_\varepsilon^{opt *}(r') 
\end{equation}

\noindent After this substitution $(M^2_w)_{val} \to 0$, and
$q^2$ tends to a finite limit when $w \to 0$. (iii) The
substitution (16) has no practical importance when the realistic
optical--model parameters are used. In this case inequality
$\Gamma^\downarrow_{sp} \gg \Gamma^\uparrow_{sp}$ is fulfilled
($\Gamma^\uparrow_{sp}$ and $\Gamma^\downarrow_{sp}$ are the
single--particle escape and spreading widths, respectively), and
as a result, the last term in eq.(16) can be neglected.

The calculations of $q^2$ show that this quantity is practically
independent (with an accuracy of $20\%$) of $w$ and mass number
$A$, but it is markedly dependent on the choice of the
formfactor $f(r)$. When the surface--type absorption is chosen
($f(r)=4a_0{df_{WS}(r,R,a_0)/{dr}}$) $q^2\approx 0.7 $.  In the
case of volume--type absorption the value of $q^2$ is less by
the factor of $3.0$.  The difference is explained by the fact
that single--pole approximation for both the single particle
Green function and the potential-scattering matrix is valid when
the considered nucleus is close to the relevant shape resonance,
but none of the nuclei can belong simultaneously to the
$s_{1/2}$ and $p_{1/2}$ shape resonances. Therefore, the
single--particle weak matrix element in the representation
$q^2=(M^2_w)_{sp}/(\Gamma^{\uparrow 0}_{s_{1/2}})_{sp}
(\Gamma^{\uparrow 0}_{p_{1/2}})_{sp}$ \cite{ur2} is not
well-defined quantity.

For estimation of the upper limit of the valence-mechanism
contribution to $M_w$ according to eqs.(15) we use: (i) the
surface-type formfactor of the imaginary part of the optical
potential; (ii) the strength of the neutron-nucleus weak
interaction $\xi_n=3.3$; (iii) the experimental values of
neutron widths $\Gamma^0_{s_{1/2},p_{1/2}}=
S^0_{s_{1/2},p_{1/2}}d_{s,p}$. The calculated ratios
$r=(M_w)_{val}/M_w$ are given in the last column of the Table I. 
As a rule these ratios are about $10^{-3}$.

{\bf 5.} In the present work we extend the temperature version
of the shell-optical-model approach for describing the
low-energy compound-to-compound transitions induced by external
single-particle fields. On this base we evaluate the
root-mean-squared matrix elements $M_w$ for the weak mixing of
$s_{1/2}$ and $p_{1/2}$ neutron resonances.  The unified
description for the electromagnetic $c-c'$ transitions and the
P-odd mixing of nuclear compound states allows us to evaluate
the intensity of this mixing without the use of free parameters.
It is also quantitatively shown that the valence-mechanism
contribution to $M_w$ is small.  The satisfactory description of
all known experimental data on $M_w$ for intermediate and heavy
mass nuclei has been obtained within the framework of the given
approach.

Authors are grateful to P.A.Krupchitsky, L.B.Pikelner for the
discussions about experimental data and to S.E.Muraviev for
valuable remarks.

This work was supported in part by Grant No.MQ2000 from the
International Science Foundation, by Grant No.MQ2300 from the
International Science Foundation and Russian Government and by
Grant No.95-02-05917-a from the Russian Foundation for Basic
Research. The authors are also grateful to the International
Soros Science Education Program for support (respectively,
Grants a1459 and 444p from the Open Society Institute, New
York).
\newpage

\noindent
TABLE I. Calculated matrix elements $M_w$ in comparison with
experimental values of $M_w^{exp}$ taken from refs.
\cite{bow,shar,krup}. Calculated ratios $r=(M_w)_{val}/M_w$ are also
given.

\begin{center}
\begin{tabular}{|c|c|c|c|c|c|c|c|c|}
\hline
 &  &  &  &  &  &  &  & \\
$compound$&$d_s,$&$\alpha,$&$s^{n}_w,$&$s^{p}_w,$&$M_w^{(I)},$& 
$M_w^{(II)},$ & $M_w^{exp},$ & $r,$ \\
$nucleus$&$eV$&$MeV^{-1}$&$neV$&$neV$&$meV$&$meV$&$meV$
&$10^{-3}$ \\
\hline
 &  &  &  &  &  &  &  & \\
$^{57}Fe$ & $1.7\cdot 10^4$ & $0.10$ & $31$ & $14$ & $54$ & $92$ 
& $46^{+15}_{-12}$\ [3]& $19$ \\
 &  &  &  &  &  &  &  & \\
$^{82}Br$ & $94$ & $0.18$ & $17$ & $43$ & $6.7$ & $7.8$ 
& $3.0 \pm 0.5$\ [3] & \\
 &  &  &  &  &  &  & $2.6 \pm 0.1$\ [3] & \\
$^{108}Ag$ & $16$ & $0.07$ & $25$ & $6.8$ & $1.2$&$2.3$&$^{*)}$ 
& $1.0$ \\
 &  &  &  &  &  &  &  & \\
$^{110}Ag$ & $14$ & $0.10$ & $31$ & $8.4$ & $1.2$ &$2.5$&$^{*)}$ 
& $0.9$ \\
 &  &  &  &  &  &  &  & \\
$^{112}Cd$&$20$&$0.05$&$18$&$4.3$&$1.1$&$2.2$&
$1.6^{+0.8}_{-0.5}\ [3]$ & $1.6$ \\
 &  &  &  &  &  &  & $2.6 \pm 0.4$\ [3] & \\
$^{114}Cd$ & $21$ & $0.11$ & $33$ & $8.0$ & $1.5$ & $3.1$
& $2.0^{+1.6}_{-0.9}$\ [2] & $0.7$ \\
 &  &  &  &  &  &  & $0.4 \pm 0.1$\ [3] & \\
 &  &  &  &  &  &  & $0.84 \pm 0.23$\ [3] & \\
$^{116}In$ & $9$ & $0.05$ & $15$ & $3.0$ &$0.6$&$1.4$
&$0.59^{+0.25}_{-0.15}$\ [2]&$0.8$ \\
 &  &  &  &  &  &  &  & \\
$^{118}Sn$ & $48$ & $0.03$ & $10$ & $2.2$ & $1.2$ & $2.5$
& $0.7 \pm 0.1$\ [3] & $1.7$\\
 &  &  &  &  &  &  & $3.7$\ [3] & \\
$^{140}La$ & $208$ & $0.04$ & $5.5$ & $11$ & $5.0$ & $6.1$ 
& $1.7 \pm 0.1$\ [3] & $2.3$\\
 &  &  &  &  &  &  & $1.3 \pm 0.1$\ [3] & \\
 &  &  &  &  &  &  & $4.0$\ \cite{pik}& \\
$^{233}Th$&$17$&$0.08$&$18$&$9.0$&$1.4$&$2.2$& 
$1.39^{+0.55}_{-0.38}$\ [1] & $1.1$ \\
 &  &  &  &  &  &  &  & \\
$^{239}U$& $21$ &$0.08$&$19$&$8.0$&$1.5$& $2.5$ 
& $0.56^{+0.41}_{-0.20}$ \ [1]& $1.6$\\
 &  &  &  &  &  &  &  & \\
\hline 
\end{tabular}
\end{center}

$*)$ The data are expected \cite{shar}

\end{document}